# WTO GPA and Sustainable Procurement as Tools for Transitioning to a Circular Economy

Sareesh Rawat

The George Washington University Law School

# Table of Contents

# 1 ABSTRACT

We live in an age of consumption with an ever-increasing demand of already scarce resources and equally fast growing problems of waste generation and climate change. To tackle these difficult issues, we must learn from mother nature. Just like waste does not exist in nature, we must strive to create circular ecosystems where waste is minimized and energy is conserved. This paper focuses on how public procurement can help us transition to a more circular economy, while navigating international trade laws that govern it. The paper is divided into three parts. Part I depicts through examples; how circular economy principles can be applied in different stages of the procurement lifecycle to achieve sustainable development objectives. Part II discusses the major non-discrimination and transparency principles of GATT and the Revised GPA and the trade issues arising out a potential conflict between circular procurement practices and these principles' obligations. Part III contains a case study on Biofuels and the current production, innovation and international trade challenges the biofuel market faces. Finally, the paper concludes with a suggestion to use circular procurement practices to promote innovation in biofuel production using food waste as a feedstock at large DoD facilities.

# 2 INTRODUCTION

Public procurement plays a key role in implementing social, political and economic policies. Acquisition of products and services can be used to promote various governmental interests, while balancing various economic, social and environmental factors. According to an OECD report, public procurement markets account for 10-15% of GDP in developed countries.[1] This level of spending allows governments to utilize public procurement as an effective policy tool to influence outcomes in domestic markets and in international trade. With government procurement accounting for such a large part of a country's economy, practicing sustainable procurement is critical. In the United States, there have been several procurement programs and policies adopted since the 1970s to promote sustainable procurement. Similarly, various European Union directives require purchase of environmentally preferable products.

---

[1] OECD, *PRINCIPLES FOR INTEGRITY IN PUBLIC PROCUREMENT* 9 (2009).

Government procurement inspired by the principles of circular economy has recently been recognized as an effective way to achieve sustainability goals. While there is no standardized definition of circular procurement, it can be broadly defined as a restorative and regenerative system that aims to keep products, components and materials at their highest possible utility and value at all stages of the procurement lifecycle.[2] Since sustainable procurement policies can vary at national and sub-national level, it is critical to define sustainability objectives that are compatible with primary procurement principles such as value for money. Ideally a circular approach to procurement succeeds when the government meets its needs in a way that achieves value for money throughout the lifecycle, for the government and for the wider society, while minimizing material loss and environmental impact.[3] In terms of countries' international trade obligations to each other, it is also important that sustainable procurement be practiced without violating essential international trade principles such as non-discrimination.

---

[2] ELLEN MACARTHUR FOUND., DELIVERING THE CIRCULAR ECONOMY: A TOOLKIT FOR POLICYMAKERS 19 (2015), https://www.ellenmacarthurfoundation.org/assets/downloads/publications/EllenMacArthurFoundation_PolicymakerToolkit.pdf.
[3] *Id.* at 170.

### 3  PART I: APPLYING CIRCULAR ECONOMY PRINCIPLES TO ACHIEVE SUSTAINABLE PROCUREMENT OBJECTIVES

Circular economy has three main principles that focus on closing the loops of product lifecycle by saving values at various stages. These include: (1) designing out waste and pollution; (2) keeping products and materials in use; and (3) regenerating natural systems.[4] It is a contrasting approach to the more linear model of consumption that entails "take, make, dispose." Public procurement can play an important role in transitioning to circular economy.

### 3.1  Outlining the Need to Switch from a Linear to Circular Model.

The current linear approach to procurement has three primary stages, each involving a cost to the procuring entity. The first stage involves the purchase of the product, which involves the purchase price of the product. The second stage in a linear approach encompasses the costs of making use of the product and the cost of meeting relevant regulations. Whereas, the third and final stage of the linear procurement lifecycles involves the costs of disposing the product consistent with the relevant regulations.[5] Along with economic costs, this procurement model encourages mass production, consumption and disposal, which directly contributes to global environmental problems such as climate change, air pollution, resource depletion and waste disposal.

Considering these issues, a shift to a more circular concept of consumption, such as the ReSOLVE[6] framework is needed. Circular economy involves using resources more efficiently through closing, extending and narrowing material loops.[7] Major benefits of circular economy include reduced burden on primary natural resources, reduced supply risk, reduced environmental problems, and new economic and job opportunities.[8] Governments can reap these benefits by modeling their procurement systems to encourage reduction or elimination of waste at different stages of the procurement lifecycle. While the lifecycle of a complicated procurement can be divided into several stages depending on the objectives of the procuring entity, a typical procurement lifecycle can be broadly divided into three main stages: (i) Procurement Planning and Sourcing; (ii) Contract Award and Management; (iii) Contract Closeout and Disposal.

---

[4] *Id*. at 19.
[5] *Id*. at 13-15.
[6] Id. at 21.
[7] *Id*.

The first stage involves the identification of a need and development of a procurement strategy. It focuses on detailing specifications and pre-award negotiations with the vendors, as applicable. The contract award and management stage involve the decision to award to a contractor usually based on 'best value' principles, vendor's responsiveness to the specifications and its responsibility as a business. It also includes post-award contract management and the performance of the contract. The final stage of the procurement lifecycle includes contract closeout and disposal of material as applicable, signaling the end of the contract. There are opportunities to apply various circular economy strategies during each of these three stages.

## 3.2 Applying the ReSOLVE Framework to Procurement Stages by Leveraging Lessons from Successful Circular Implementations.

The Ellen MacArthur foundation has identified six action areas for businesses and countries to shift to a more circular model of consumption.[9] Also known as the "ReSOLVE" framework, these action areas provide useful strategies, which can be implemented in government procurement models to reduce the wastefulness of the linear "procure, use, dispose" model. In order to move to a circular model of procurement, it is important to identify the specific stage(s) of the procurement lifecycle where strategies from the six action areas of the ReSOLVE framework can be most effectively implemented. It is also useful to study project examples where circular economy benefits were achieved through purposeful procurement practices.

---

[8] *Id.*
[9] *Id.* at 21.

These key principles encourage: (i) preserving and enhancing non-renewable and natural stocks through control and balance of renewable resource flows; (ii) optimization of resource yields by creating circular loops for products so they can be used at their highest utility at different stages of their lifecycle; and (iii) identifying and designing out negative externalities such as pollution and toxins that cause issues such as negative health effects and climate change.[10]

### 3.2.1  Action Area 1: REGENERATE

The first action area focuses on regeneration of exhaustible resources and eco-systems. It involves a shift to renewable energy and materials, and aims to reclaim, retain, and restore the health of the ecosystem.[11] Recovered biological resources are also returned to the biosphere.[12] Strategies from this action area can be applied in the procurement planning and contract closeout stages. The City of Vaasa in Finland successfully applied these strategies in their public transportation program involving busses that run on locally produced bio-gas.[13] Vaasa issues three separate contracts procuring the bio-gas from a local producer, busses and a service provider who committed to run the busses in the city for a period of five years.[14] Through this procurement, a new ecosystem was created that utilized local waste effectively and created an incentive for private firms to utilize bio-gas vehicles.[15] The busses led to an expected savings of 1,000 tons of $CO_2$ emissions per year.[16]

---

[10] *Id.*
[11] *Id.* at 22.
[12] *Id.*
[13] KATRINA ALHOLA ET AL., CIRCULAR PUBLIC PROCUREMENT IN THE NORDIC COUNTRIES 39-41 (2017).
[14] *Id.*

Effective planning and a thorough market research study in the first stage of the procurement was key to the success of this program. Vaasa's procurement of the busses led to a shift of risk from the service provider to the procurer, thus incentivizing the service provider. City's capacity to locally produce and utilize bio-gas effectively was also researched prior to the procurements.[17] This circular ecosystem can be further promoted through City's private garbage collection contracts requiring separate streams for organic waste, which has the potential of generating cost savings in the long run.

### 3.2.2 Action Area 2: SHARE

The second action area involves sharing of assets. It aims to maximize utilization of products through sharing between users.[18] The sharing can be done through re-using the products for their entire technical lifetime i.e., using second-hand products or by peer-to-peer sharing.[19] This area also focuses on prolonging product lifetime by maintenance, repair and durable design.[20] Sharing between users primarily applies to the contract management phase of public procurement, where the agencies can utilize procured products more efficiently through sharing.

Examples can include vehicle sharing and leasing facilities when not in use. An agency within the City of Bremen, Germany reduced $CO_2$ emissions from business related travel by replacing its fleet of 11 vehicles with a subscription to a car sharing service.[21] Through the implementation of this program, the agency claimed to have reduced the average $CO_2$ emissions from 141 g/km to 102 g/km, while also saving 9.6 cents/km amounting to over 21% cost savings.[22] Success of this program was based upon sharing of existing resources applied in the furtherance of contract performance objectives. An ancillary benefit of the program to the procurement system was higher cost transparency for car usage.

---

[15] *Id.*
[16] *Id.*
[17] *Id.*
[18] Ellen Macarthur Found., *supra note 7*, at 23.
[19] *Id.*
[20] *Id.*

### 3.2.3  Action Area 3: OPTIMIZE

The third action area focuses on increasing the performance and efficiency of the product, as well as removing waste from the product lifecycle without changes to the actual product.[23] Also known as the "optimize" phase of the ReSOLVE framework, this area encourages use of innovation and leveraging information technology such as big data, automation, remote sensing and steering.[24] Governments can implement strategies in this focus area for improving all stages of procurement through increased insight that results in improved policymaking.

Zurich's procurement of an Optimized Output Management Service to replace its investment in buying or renting multi-functional printing devices is an example of successful implementation of optimization strategies in public procurement.[25] The vendor was required to provide a "FollowMe-Printing-Solution" that held all output in a large encrypted storage, only printing after a user physically logged into a printer.[26] The program had several benefits including avoidance of careless printing, avoidance of toner waste and timely replenishment by professionals, a whopping 34% energy savings and reduction of printers and models.[27] Specific technical criteria aimed at achieving these benefits were planned and included in the requirements of the solicitation and meeting of these requirements was prioritized even more than the cost of the solution in the resultant award. The contractor also had to pay for and take over the existing printing devices, properly disposing the older devices and re-using the newer devices.[28] This environmental and cost benefits of this solution were achieved due to effective implementation of circular procurement strategies in all three phases of the procurement lifecycle.

---

[21] EU INTELLIGENT ENERGY EUROPE PROGRAMME, REPORT ON CLEAN FLEETS, INCREASING EFFICIENCY OF ADMINISTRATION'S FLEET MANAGEMENT: CAR-SHARING IN BREMEN 2 (2015), http://www.clean-fleets.eu/fileadmin/files/documents/Publications/case_studies/Clean_Fleets_case_study_-_Bremen_Car-Sharing_integration.pdf.

[22] *Id*. at 4.

[23] Ellen Macarthur Found., *supra note 7*, at 23.

[24] *Id*.

[25] EUROPEAN COMMISSION, REPORT ON PURCHASING COPY, PRINTING AND SCANNING SERVICES IN ZURICH 1 (2015), http://ec.europa.eu/environment/gpp/pdf/news_alert/Issue53_Case_Study108_zurich_output_management.pdf.

### 3.2.4  Action Area 4: LOOP

The fourth action area of the framework involves the concept of looping which is most symbolic of circular procurement ideals. It stresses on keeping product components and materials in closed loops, while prioritizing inner loops.[29] Non-renewable materials and resources can be looped through remanufacturing products or components and recycling materials. On the other hand, renewable materials looped through processes such as anaerobic digestion and extraction of biochemical from organic waste.[30] In the procurement lifecycle, this action area can be implemented most effectively in the project closeout and disposal stage. Benefits of this strategy can also be achieved through proper procurement planning by looking for looping opportunities for products already in the procurement lifecycle.

---

[26] *Id.*

[27] *Id.* at 3.

[28] *Id.*

[29] Ellen Macarthur Found., *supra note 7*, at 23.

[30] *Id.*

Dutch Ministry of Defense's procurement of towels and overalls with at least 10% re-cycled textiles fibers, when combined with the requirement that the Ministry's surplus and used clothing is no longer burnt but re-cycled is an implementation example of looping strategies.[31] As a result as high as 36% of used textiles is incorporated into the new textiles successfully maintaining the utility value of the used textile post-consumption through looping.[32]

### 3.2.5  Action Area 5: VIRTUALIZE

The penultimate action area of the ReSOLVE framework dematerializes resource use through a virtual delivery of utility, either directly or indirectly.[33] It aims at cutting out intermediary products and materials that are no longer necessary for the delivery of a utility. An example of direct virtualization includes discontinuation of CDs, DVDs and other similar storage devices for books, music and movies.[34] Online shopping through websites such as Amazon is an example of indirect virtualization.[35] Virtualization strategies can be implemented in the procurement planning, sourcing, contract award and contract management phases of the procurement lifecycle. Increased use of e-procurement methods and use of government portals such as *GSA Advantage* represent examples of successful implementation. Ancillary benefits of e-procurements include increased transparency and accountability.

---

[31] DUTCH MINISTRY OF DEFENSE, REPORT ON WORKWEAR PROCUREMENT 1 (2017), https://www.pianoo.nl/sites/default/files/documents/documents/rebusfactsheet15-kledingdefensie-engels-juni2017.pdf.
[32] *Id*. at 2.
[33] Ellen Macarthur Found., *supra note 7*, at 23.
[34] *Id*.
[35] *Id*.

### 3.2.6  Action Area 6: EXCHANGE

The sixth and final action area involves an exchange of old non-renewable materials with advanced materials and effectively utilizing new technologies to choose new products and services. This can be most effective in the procurement planning and project closeout stages of the procurement lifecycle. Zurich's procurement of the Optimized Output Management Solution as an alternative to owning and leasing printers described above is an example of successful implementation.[36] There a new technology was leveraged and a new service was procured instead to achieve the end-product. Another example is Helsinki's procurement of pilot projects to evaluate methods of treating and using digested sewage sludge from a waste treatment center.[37] These pilot projects encourage innovative solutions focusing on nutrient and material recycling rather than energy savings.[38]

As demonstrated through examples above, Ellen MacArthur foundation's ReSOLVE framework to circular economy can promote efficient circular loops in all stages of public procurement that benefit the environment and create cost-savings and other sustainability benefits in both short and long term. Benefits from circular procurement can be maximized through innovative and strategic thinking in all phases, but particularly in the planning phase of the procurement. Private firms and government contractors can be incentivized to participate by reducing risk or improved pricing for coming up with innovative solutions that promote circular loops.

As the examples in PART I demonstrate, there are multiple approaches to achieve greater circularity in public procurement. The United Nations Environment Programme (UNEP) in a report created as part of a platform to accelerate circular economy through sustainable procurement, categorizes these multiple approaches into two pillars.[39] First is to promote circular supply chains by procuring more circular products and materials.[40] Secondly to promote new business models based on innovative and resourceful solutions.[41]

---

[36] European Commission, *supra note 30*, at 1. Katrina Alhola
[37] Katrina Alhola, *supra note 18*, at 36.
[38] *See Id.* (stating that procurer was focused on developing innovative methods to produce fertilizers or biochar and conducted a market survey for pilot projects before implementation).

As a strategy to achieve the first pillar, the UNEP report suggests using circular procurement criteria in bid specifications.[42] This is consistent with our analysis that circular economy can be most effectively accelerated through the first stage in procurement, as they can be included within the technical specifications and requirements of the bid request. Once these are part of the evaluation criteria, it gives the contractors an incentive to promote circular economy to win government business.

## 4 PART II: BALANCING CIRCULAR PROCUREMENT & INTERNATIONAL TRADE OBLIGATIONS

While including sustainability procurement requirements in a solicitation or applying them at a later phase in the procurement may accelerate circular procurement, including these requirements as technical criteria can have trade implications as it relates to the non-discrimination requirements of the WTO General Agreement on Tariffs and Trade (GATT) and more specifically of the Government Procurement Agreement (GPA). Part II of this paper provides an overview of relevant GATT and GPA principles and their impact on trade between member states as it relates to government procurement. This is done by providing a brief overview of the relevant provisions and their scope in government procurement. A case study on bio-fuel is then provided in Part III as an example to demonstrate the trade related complexities that may arise when governments include criteria promoting circular economy in their purchases.

---

[39] UNEP, SHAPING THE FUTURE OF ENVIRONMENT AND NATURAL RESOURCE SECURITY BUILDING CIRCULARITY INTO ECONOMIES THROUGH SUSTAINABLE PROCUREMENT (INITIAL FINDINGS) 5 (2018), http://www3.weforum.org/docs/Environment_Team/40049_Shaping_Future_Environment_Natural_Resource_Security_report_2018.pdf.
[40] *Id*.
[41] *Id*.
[42] *Id*. at 11.

## 4.1 Assessing Primary and Secondary Procurement Objectives to Define "Best Value."

Any government procurement is said to have two objectives, which can be at times in conflict with one another.[43] A primary objective which can be described as fulfilling governmental requirements while achieving economic efficiency as public funds are spent.[44] This has been described as the "best value" or "value for money" concept as it relates to the tax payer funds.[45] In other words it means getting the best possible goods "among a set of similarly priced category of same or similar goods."[46] The primary objective is best achieved when the procurement system has adequate competition between vendors, transparency and accountability.

Secondary objectives of a government procurement are said to include non-economic objectives such as social, political and environmental objectives that a government would like to achieve to further its national policy.[47] A government may have different approaches to meet these objectives through procurement. This could be done in the form of subsidies, set-asides or preferences in the evaluation criteria for meeting these objectives.[48] Trading in government procurement is complex as the governments' role in meeting these domestic secondary objectives is tied directly to the country's sovereignty. However, instruments such as the GPA aim to promote trade in government procurement, especially in areas not directly tied to issues of sovereign governmental functions. The question thus arises, how can governments achieve secondary objectives of domestic procurement, such as those that promote a circular economy for the benefit of the environment, without violating important non-discrimination principles of the GPA.

---

[43] GARBA I. MALUMFASHI, "GREEN" PUBLIC PROCUREMENT POLICIES, CLIMATE CHANGE MITIGATION AND INTERNATIONAL TRADE REGULATION: AN ASSESSMENT OF THE WTO AGREEMENT ON GOVERNMENT PROCUREMENT 76 (1995).
[44] *Id.*
[45] *Id.*
[46] *Id.* at 77.
[47] *Id.*
[48] See *Id.* at 66 (stating other covert discriminatory practices).

### 4.1.1 Government Procurement Exception in the GATT & Adoption of the GPA

Due to the secondary objectives of government procurement being motivated by non-economic reasons, governments were reluctant to allow government procurement as an area that was subject to international competition.[49] Thus government procurement was not subject to one of two main non-discrimination requirements of the GATT Multilateral Trading System.[50] The National Treatment requirement under GATT Article III was not applicable to government procurement since the initial negotiations in 1946.[51] Specifically, Article III 8(a) of the GATT states that the "provisions of this Article shall not apply to laws, regulations or requirements governing the procurement by governmental agencies of products purchased for governmental purposes and not with a view to commercial resale or with a view to use in the production of goods for commercial sale."[52] Even though Article III only deals with the National Treatment principle of the GATT, it also effects the other non-discrimination principle of 'Most Favored Nation' as these two principles are inextricably related.

---

[49] Malumfashi, *supra note 44*, at 88.
[50] *Id*.
[51] *Id*. at 89.
[52] General Agreement on Tariffs and Trade, art III 8(a), Oct. 30, 1947, 61 Stat. A-11, 55 U.N.T.S. 194 [hereinafter GATT].

This practice of allowing governments to practice discriminatory government procurement until the Tokyo Round Government Procurement Code came into force in 1979, after being negotiated on a plurilateral basis.[53] This code extended the 'Most Favoured Nation' and 'National Treatment' obligations of international trade, along with transparency rules at all stages of the procurement.[54] However, these obligations were conditional upon reciprocal promises between the signatories, thus limiting it in coverage.[55] As applied in 1996 after two revisions, this code was expanded as the 'Revised Government Procurement Agreement' to include a broader coverage of governmental entities within the signatories, as well as a broader scope to include services and certain other areas not previously covered.[56] The latest revision of the GPA as revised GPA was adopted in 2012 and it entered into force on April 6, 2014.

### 4.1.2  Coverage of the Revised GPA

The World Trade Organization (WTO) describes the revised GPA as the pre-eminent international instrument regulating the conduct of trade in government procurement markets, that aims to ensure fair, transparent, and non-discriminatory conditions for purchases of goods, services and construction services by the public entities covered by the Agreement.[57] Currently a total of 47[58] WTO members are covered by the GPA and another 31[59] WTO members and four[60] international organizations participate in the Committee as observes.[61]

---

[53] Malumfashi, *supra note 44*, at 90.
[54] *Id*.
[55] *Id*.
[56] *Id*. at 97.
[57] *See* Revised Agreement on Government Procurement, Mar. 30, 2012, Marrakesh Agreement Establishing the World Trade Organization, Art. II, 1915 U.N.T.S. 103 [hereinafter 2012 GPA].

As a "closed" plurilateral agreement with limited coverage, the GPA is applicable to certain types of public procurements between member countries. Appendix I to the GPA contains the coverage schedules that list for each member country: (i) the procuring entities covered under the agreement; (ii) the goods, services and construction services covered; (iii) threshold values for the procurement activities to be covered under the agreement; and (iv) exceptions to the coverage under the agreement.[62] Coverage schedules for every GPA member country has seven annexes and can be found on the e-GPA portal.[63]

---

[58] *See Members and Observers 2012 GPA*, https://www.wto.org/english/thewto_e/whatis_e/tif_e/org6_e.htm (last visited Apr. 14, 2019) (including member countries Armenia, Canada, European Union (including its 28 member states), Hong Kong - China, Iceland, Israel, Japan, Korea, Liechtenstein, Montenegro, Netherlands with regard to Aruba, Norway, New Zealand, Singapore, Switzerland, Chinese Taipei, United States).

[59] *Id.* (stating that observer WTO countries include Albania, Argentina, Australia, Bahrain, Cameroon, Chili, China, Colombia, Costa Rica, Georgia, India, Indonesia, Jordan, Kyrgyz Republic, Malaysia, Moldova, Mongolia, Oman, Pakistan, Panama, Russian Federation, Saudi Arabia, Sri Lanka, Tajikistan, Thailand, The former Yugoslav Republic of Macedonia, Turkey, Ukraine, and Viet Nam).

[60] *See International Intergovernmental Organizations Granted Observer Status to WTO Bodies*, https://www.wto.org/english/thewto_e/igo_obs_e.htm (last visited Apr. 14, 2019) (Stating that the four international organizations with observer status to the GPA are: International Monetary Fund (IMF), the Organization for Economic Cooperation and Development (OECD), the United Nations Conference on Trade and Development, and the International Trade Centre).

[61] Out of the observer countries, Albania, Australia, China, Georgia, Jordan, Moldova, Mongolia, Oman, The former Yugoslav Republic of Macedonia, Turkey and Ukraine are either negotiating accession to the GPA or have initiation of accession negotiations pending).

[62] *See* 2012 GPA, App'x I.

[63] *See e-GPA Portal,* https://e-gpa.wto.org/ (last visited Apr. 14, 2019) (Stating that schedule of each party has the following seven annexes: Annex 1: Central government entities; Annex 2: sub-central government entities; Annex 3: other entities; Annex 4: goods; Annex 5: services; Annex 6: construction services; Annex 7: general notes).

The GPA is estimated to provide market access to procurement markets currently estimated at $1.7 Trillion annually.[64] The latest revision of the GPA updated the agreement's text to provide the members with wider access to government procurement markets and encourages the use e-procurement.[65] But most importantly in regard to promoting circular economy in public procurement, the revised GPA provides a more explicit recognition of the right of procuring entities to promote environmental values and sustainability.[66] Article X(6) of the GPA allows parties to include technical specifications in their procurements to promote their environmental interests. It states, "a Party, including its procuring entities, may, in accordance with this Article, prepare, adopt or apply technical specifications to promote the conservation of natural resources or protect the environment."[67] Later, Part III of this paper discusses the practical ways this article of the revised GPA could be applied with the help of a case study.

### 4.1.3 Main Principles of the Revised GPA

The revised GPA is primarily based on the principles of non-discrimination for covered procurements, transparency of procurement, and fairness in procedure.

#### 4.1.3.1 Non-Discrimination

The non-discrimination principles of National Treatment and Most Favored Nation treatment for covered procurements can be found in Article IV(1) of the 2012 GPA. It states:

> "With respect to any measure regarding covered procurement, each Party, including its procuring entities, shall accord immediately and unconditionally to the goods and services of any other Party and to the suppliers of any other Party offering the goods or services of any Party, treatment no less favourable than the treatment the Party, including its procuring entities, accords to:
>> a. domestic goods, services and suppliers; and

---

[64] *See What is the GPA?*, https://www.wto.org/english/tratop_e/gproc_e/gp_gpa_e.htm (last visited Apr. 14, 2019).
[65] *See* 2012 GPA, Art. IV(3).
[66] 2012 GPA Art. X(6).
[67] *Id.*

b.   goods, services and suppliers of any other Party."[68]

As in the GATT, the national treatment principle contained in Article IV(1)(a) prevents members

from discriminating against covered products or services of foreign countries that are also parties

to the GPA. On the other hand, the most favored nation principle included in Article IV(1)(b)

compels member countries to treat covered products and services of other member countries as equal. It is important to note that the GPA only extends the market access benefits provided by its non-discrimination clause to the parties of the agreement, making it a "closed" agreement, benefits of which are not available to WTO members as a whole.

### 4.1.3.2  Transparency & Procedural Fairness

Through its transparency provisions, the GPA has established certain minimum standards for

covered procurements. These include publication of procurement legislation and new

opportunities covered under the agreement to ensure adequate notice to potential suppliers.[69]

Specific transparency provisions also require member states to conduct covered procurement in a

manner that avoids conflict of interest and prevents corrupt practices.[70] Article IV(4) of the GPA

states:

> "A procuring entity shall conduct covered procurement in a transparent and
> impartial manner that:
>> a.   is consistent with this Agreement, using methods such as open
>> tendering, selective tendering and limited tendering;
>> b.   avoids conflicts of interest; and
>> c.   prevents corrupt practices."

Award Criteria in Article XIII(4)(a) of the GPA allows the award to lowest tender or "most advantageous"

which can potentially include environmental and circular economy benefits in performing the contract.

---

[68] 2012 GPA Art. IV(1).
[69] *See* 2012 GPA Art. VII.
[70] 2012 GPA Art. IV(4).

It is interesting to note that Award Criteria in Article XIII(4)(a) of the GPA allows the award to lowest bidder or "most advantageous bidder" which means that environmental and circular economy factors can be considered along with economic factor in deciding which bidder represents the highest overall value to the government. GPA also includes fairness provisions such as the independent domestic review of complaints to ensure that suppliers from member states have a dispute settlement mechanism for disputes arising from fulfilling procurement requirements of other member states.[71]

## 5 PART III: BIOFUELS – INTERNATIONAL MARKET LANDSCAPE, CHALLENGES & USING GOVERNMENT PROCUREMENT TO ACCELERATE A CIRCULAR SOLUTION

Biofuel is used as an case-study, as it is seen as a product that can help countries in achieving their policy goals such as energy security and mitigating climate change.[72] Biofuels when mixed with conventional fuels, help reduce emissions such as sulfur particulates, carbon-monoxide and hydrocarbons. Procurement of bio-fuel instead of conventional fuel by government agencies can help promote circular economy. Procurement by the defense sector in particular can have a substantial impact due to it being the largest consumer of energy.

### 5.1 Governmental Measures to Protect & Promote Biofuel

Governments use several measures to promote the use of biofuel at different stages of production and use chain.[73] Mandates for blending of biofuels with petrol or diesel have been issued by all major biofuel producing countries.[74] Reductions in tax and financial assistance to promote the research and development of biofuels is also common.[75] Some measures to promote biofuel include targets that require a particular percentage of biofuel to be included the nation's total fuel supply,[76] tax-credits for companies engages in biofuel production, along with other incentives and operating grants.[77]

---

[71] *See* 2012 GPA Art. XVIII.
[72] Marsha A. Echols, Int'l Ctr. for Trade & Sustainable Dev., *Biofuels Certification and the Law of the World Trade Organization,* at ix-x, (Issue Paper No. 19) (June 1, 2009).
[73] Toni Harmer, Int'l Ctr. for Trade & Sustainable Dev., *Biofuels Subsidies and the Law of the WTO*, at 16, Issue Paper No. 20 (June 1, 2009).
[74] *Id.*

Government support for biofuel in the production stage include loans, loan guarantees, and tax incentives for infrastructure costs such as accelerated depreciation.[78] Grants for biofuel production include infrastructure capital grants and business planning and market development assistance.[79] Government also utilize general subsidy programs for feedstock including sugar and corn, and subsidies for indirect inputs such as fertilizers, water and seeds.[80] Governments also promote the production of specific biofuel crops for energy use, including direct payments for land used to grow such energy crops.[81] Support for distribution and use of biofuel includes fuel-tax reductions to compensate customers for higher cost of biofuel production when compared to conventional fossil fuels, thereby encouraging the use of alternative fuels.[82] Governments also provide incentives for purchase of vehicles that can run on bio-fuel either through direct rebates or through reduction in taxes.[83] Finally assistance is also provided with the costs of biofuel refueling and storage infrastructure costs.[84]

---

[75] *Id.*
[76] *Id.*
[77] *Id.*
[78] *Id.*
[79] *Id.*
[80] *Id.*
[81] *Id.*
[82] *Id.*
[83] *Id.*
[84] *Id.*

**5.2 International Trade in Biofuels**

Biofuel markets are affected by factors such as sustainability criteria, fuel quality standards and import tariffs on ethanol and biodiesel.[85] According to the OECD-FAO Agricultural Outlook 2016-2025, major biofuel producers such as United States, the European Union, Brazil and Argentina, all have frameworks of domestic policies and expanding mandates for biofuel.[86]

In the United States, Biofuel being a "bio-based" product falls under farm laws.[87] The Energy Independence and Security Act established the Renewable Fuel Standard Programme (RFS2), which set four quantitative mandates until the year 2022.[88] In the European Union the 2009 Renewable Energy Directive[89] and the Fuel Quality Directive[90] determine the policy framework for biofuels.[91] This framework was amended in 2015 via the Indirect Land Use Changes (ILUC) directive, which introduced a 7% cap on renewable energy from food and feed crops in the transport sector.[92]

**5.2.1 Biofuel Mandates as Barriers to Trade**

There are several barriers to trade that effect biofuel. Most major biofuel producing countries have mandates that cause Biofuel to stay domestic. In Brazil, the domestic demand for biofuels is sustained through a mandatory 27% blending requirement of anhydrous ethanol mandatory blending requirement for gasohol.[93] There is also a differentiated taxation system that is favorable to hydrous ethanol.[94] In Argentina, the biodiesel mandate is expected to increase from 10% in volume terms in 2017 to 14% by 2025, despite a reduction in Indonesian biodiesel import.[95] Argentina is a good example of a country where the government support policies have led to the strong growth of the domestic biodiesel sector and the utilization of domestic palm oil resources to replace foreign imports.[96]

---

[85] OECD-FAO, AGRICULTURAL OUTLOOK 2016-2025 116 (2016), http://www.fao.org/3/a-BO103e.pdf.
[86] *Id.* at Biofuels-1.
[87] *Id.* at Biofuels-2.
[88] *Id*. at Biofuels-1 (stating that "the total and advanced mandates that require fuels to achieve respectively at least a 20% and a 50% GHG reduction as well as the biodiesel and the cellulosic mandates that are nested within the advanced mandate").
[89] *Id*. (stating "that renewable fuels (including non-liquids) should increase to 10% of total transport fuel use by 2020 on an energy equivalent basis").
[90] *Id*. (requiring "fuel producers to reduce the GHG intensity of transport fuels by 2020").
[91] *Id*.
[92] *Id*.

For other relatively minor biofuel markets around the world, effective government policy support along with international price trends are the major factors contributing to development.[97] For instance, the Indian government is expected to enforce an E10[98] mandate which will spur the demand for ethanol.[99] This demand however, is expected to be covered by domestic ethanol production from molasses which will compensate for the high sugar cane prices faced by sugar mills.[100] Due to self-enforced policies Thailand is expected to contribute to 11% of the world's ethanol production increase by 2021.[101] Thailand government is expected to utilize heavy subsidies for gasohol and biodiesel to meet its targets.[102]

[93] *Id.* (describing Gasahol as a mixture of gasoline and anhydrous ethanol and that the use of vehicles that can either run on gasohol or hydrous ethanol is prevalent in Brazil).
[94] *Id.*
[95] *Id.* (stating that Argentinian biodiesel production is heavily dependent on policies in palm oil supplying countries, especially Indonesia).
[96] *Id.*
[97] *Id.*
[98] Targay, *What is E10 Ethanol?*, https://www.targray.com/biofuels/blends/e10-ethanol (describing E10 ethanol as a low concentration biofuel blend consisting of 10 percent ethanol along with 90 percent gasoline).
[99] OECD-FAO, *supra note 90*, at Biofuels-3.
[100] *Id.*
[101] *Id.*
[102] *Id.*

### 5.2.2 Biofuel Trade Projections & Expectations

By 2020, OECD anticipates a modest increase in trade of ethanol[103], which is the most traded biofuel currently in global markets. Export of United States' ethanol is however not expected to grow too much.[104] This is despite the fact that United States is expected to remain a net exporter of ethanol made from maize feedstock, and a modest importer of sugarcane based ethanol during this period.[105] In the European Union, ethanol import requirements are predicted to reach their maximum levels of 1.9 Bln L in 2020, an increase of almost 1.5Bln L since 2015.[106] Despite being one of the largest producers of ethanol, Brazil's export of ethanol is expected to remain stable in the next few years since most of Brazil's ethanol will be used domestically.[107]

Biodiesel trade on the other hand is projected by OECD to remain stable relative to ethanol, with Argentina expected to remain as the largest exporting country.[108] United States biodiesel mandate will primarily be fulfilled using Argentinian biodiesel.[109] The European Union, however, is expected to reduce its biodiesel import demand by 2025 due to increased sustainability requirements and tariffs implemented.[110] Due to these reasons, Indonesian exports of palm based biodiesel will be marginal as there have been significant drop in volume of export since the 2012 peak.[111]

---

[103] *Id.* at Biofuels-8.
[104] *Id.*
[105] *Id.*
[106] *Id.*
[107] *Id.*
[108] *Id.*
[109] *Id.*
[110] *Id.*
[111] *Id.*

### 5.2.3 Biofuel Trade Barriers

Major impediments to biofuel trades come from high import tariffs and technical barriers to trade. Since biofuel production in most cases is currently more expensive than the price of conventional fuel, almost all of the biofuel produced in the world currently is used domestically to meet environmental mandates similar to the ones described above. Liberalization of trade in the international biofuel markets would allow a few low-cost producers (such as Brazil) to greatly expand their market share, while high-cost producers currently enjoying preferential trade agreement benefits could end up with a lower market share.[112]

Biofuels have a narrow margin of competitiveness, impacting the tradability of ethanol.[113] The consumption of biofuels in the long run can only be economically viable if marginal costs of production are equal to or lower than fossil fuels. A myriad of external factors dictate the viability of fossil fuels and biofuel economics can be inherently self-limiting.[114] For instance, ethanol from sugar in Brazil was competitive with petroleum in 2005, however higher sugar prices resulting from greater production of ethanol eroded Brazilian ethanol's margin of competitiveness.[115] Due to this very few countries besides Brazil have potential to be large exporters of ethanol or other biofuels.[116]

---

[112] *Id*. (providing examples of CBI countries currently enjoying preferential trade agreements).

[113] *See* Masami Kojima et al., Considering Trade Policies for Biofuels 57 (2007) (stating that small competitive margin explains why only a tenth of global biofuels produced and sold in the world are internationally traded).

[114] *Id*.

[115] *See Id*. (providing the example "as Brazil's ethanol production from sugarcane increases, the supply of sugar on the international markets declines and thereby raises the price of sugar. The rising price of sugar will induce sugarcane to be redirected back into sugar production and away from Ethanol").

[116] *Id*.

Security of supply of biofuels is also a key issue with trade likely to be hindered if biofuel supply poses issues similar to petroleum supply.[117] If countries with low-cost of biofuel production are concentrated in a small geographic areas, it could hinder biofuel trade due to potential insecurity of supply even if the one or two exporting countries are very well suited to biofuel production.[118]

Another important issue hindering international trade in biofuels is the classification of certain types of biofuels (such as Ethanol but not biodiesel) as agricultural goods. Agricultural goods often enjoy greater protection than industrial goods and once they receive that protection, reform is hard to achieve.[119] Trade negotiations for agricultural goods such as Ethanol fall under the Agreement of Agriculture and agricultural goods are generally excluded from coverage under the GPA.[120]

The development of biofuel markets has been strongly dependent on biofuel policy packages, the macroeconomic environment and the price of petroleum.[121] Thus, uncertainty in biofuel policy decisions in major biofuel producing countries could have detrimental effects in both short and long term. For instance, in the United States there is major uncertainty stemming from the advanced and the biodiesel mandates.[122]

---

[117] *Id*. at 58.

[118] *See Id*. (giving example of Japan as a country that is ill-suited to biofuel production but "interested in biofuel production for their GHG emission reduction benefits, worries that Brazil is now the only obvious large exporter and views reliance on one exporting country as potentially compromising security of supply").

[119] *See id*. at 62.

[120] *See* 2012 GPA Annex 1 (showing that agricultural good are excluded from the United States coverage schedule under Annex 1, Note 2).

[121] OECD-FAO, *supra note 90*, at Biofuels-8.

[122] *See Id*. (stating that while it is likely that the advanced mandate will increase over the next decade, same is not clear in regard to the biodiesel mandate).

Another factor hindering the development of biofuels is the current low prices of conventional energy and fuel prices.[123] The low prices prevent investment in research and development for advanced biofuels, such as biofuels produced from lingo-cellulosic biomass, waste or other non-agricultural feedstock.[124] This leads to the dependence of major biofuel (such as ethanol) production on agricultural crops such as sugarcane, sugar beets, maize, wheat, cassava and other starches as feedstock. While biodiesel is produced from feedstock including rapeseed oil, soybean oil, waste oil and palm oil – biodiesel is more costly than ethanol and its market is much smaller.[125] Excessive use of agricultural crops for production of biofuel is not always viable as it is limited by market forces as discussed above, and it can lead to food insecurity. Thus, any revisions to domestic policies may account for this by including more stringent sustainability criteria.[126]

### 5.2.4  Circular Procurement Solution (DoD's Procurement of Biofuel Converted from the Food Waste it produces)

Effective utilization of biofuel can have significant environmental benefits. However, there are economic impediments to development of biofuel as a major alternative to fossil fuels. Cost of feedstock comprises more than 50% of the costs of producing biofuels such as ethanol,[127] and 71.5% of the cost of producing renewable biodiesel.[128] Thus, favorable economic conditions such as high price of crude oil and low costs of feedstock are needed for biofuel to be commercially viable.

---

[123] *Id.*
[124] *Id.*
[125] MASAMI KOJIMA, supra note 118, at xiv.
[126] *Id.*
[127] *Id.* at xiv.
[128] ALEX R. MAAG ET AL., CATALYTIC HYDROTHERMAL LIQUEFACTION OF FOOD WASTE USING CEZRROX 2 (2018) (citing Department of Energy (DOE), National Renewable Energy Laboratory report stating costs of producing renewable biodiesel from biomass and municipal solid waste).

Food waste is an untapped resource considered to have great potential for generating energy. Nearly one third of all food produced gets discarded uneaten and about 70 percent of all food waste ends up in landfills, where it decomposes and adds to greenhouse gas emissions. In the United States alone, more than 15 million dry tons of food waste is generated annually, with 92% of that waste being discarded in landfills.[129] Diverting even a small portion of this waste to produce energy can significantly lessen our Carbon footprint, provide a carbon neutral source of energy and contribute to solving the global waste problem by freeing up landfill space.

While using food waste as feedstock for biofuel is not yet economically viable, scientists have made significant strides towards that goal. It is estimated that there are over 800 waste-to-energy (WTE) industrial plants around the world, along with thousands of smaller systems.[130] Research suggests that global market for WTE has the potential of growing to $29.2 billion by 2022.[131] The success of these WTE systems depend upon diversion of waste into a separate stream of organic and food waste. Some systems incentivize individuals to separate their organic and food wastes through lower trash pick-up bills.[132] These anaerobic WTE systems can be made even more efficient at reducing carbon emissions if combined with Hydrothermal Liquefaction.

---

[129] *Id.* (citing U.S. DPE, *Biofuels and Bioproducts from Wet and Gaseous Waste Streams: Challenges and Opportunities* (2017)).
[130] RODDY SCHEER & DOUG MOSS, *Food Waste to Energy*, https://www.scientificamerican.com/article/food-waste-to-energy/ (last visited Apr. 15, 2019).
[131] *Id.* (citing Navigant Research report from 2012 expecting the waste to energy market to grow from $6.2 billion to $29.2 billion in 10 years).
[132] *Id.*

In 2018, chemical engineers and researchers at Worcester Polytechnic Institute (WPI) were able to significantly improve the oil yield from waste food to energy conversion process with increased efficiencies. The project was funded by a one year Department of Energy grant worth a little over $168,000.[133] The original process called Hydrothermal Liquefaction involves wet biomass, such as food waste, being placed in a reactor and being exposed to high temperatures and pressures to breakdown hydrocarbons, producing a biofuel that is similar to crude oil.[134] However, the process is not economically viable and efficient because a significant portion of the organic compounds produced are not converted to biofuel.[135] This large portion of the organic compounds including acids and alcohols end up in the waste water produced by the reaction instead, which can be processed again to produce more usable oil.[136] But processing the waste water again leads to considerable additional costs and energy usage.[137]

As a product of their research, the team of chemical engineers and researchers at WPI were able to reduce the compound in the waste water to over 50%.[138] They did so by adding catalysts to the reaction to increase the yield of oil and make the process more efficient and economically viable.[139] The food waste feedstock used in their study comprised of a mixture representative of institutional food waste, including seven commonly disposed food items.[140] While a savings of over 50% is certainly significant, it does not yet make biofuel produced from food waste feedstock, a commercial alternative to crude oil. The WPI research in ongoing with the team experimenting with other potential catalysts to further increase the efficiency of the process.[141]

---

[133] SHARON GAUDIN, CUTTING WASTE, FOSSIL FUEL USE, AND GREENHOUSE GAS EMISSIONS BY TURNING UNUSED FOOD INTO BIOFUEL (2018), https://www.wpi.edu/news/cutting-waste-fossil-fuel-use-and-greenhouse-gas-emissions-turning-unused-food-biofuel (last visited Apr. 15, 2019).

[134] *Id.*
[135] Alex Maag, *supra note 133*, at 9.
[136] *Id.*
[137] *Id.*
[138] *Id.* at 10.
[139] *Id.* (depicting table summarizing lifetime energy yields with two types of catalysts used in experimentation); Sharon Gaudin, *supra note 138* (stating while sodium carbonate (a homogenous catalyst) did not significantly increase the yield of oil, adding a group of heterogeneous catalyst known as cerium zirconium mixed oxides did so by 40% to over 50%).

Creating biofuel using food waste feedstock is consistent with the principles of circular economy, in particular the "Loop" action area discussed as part of the ReSOLVE framework in Part I. By creating circular procurement opportunities, governments can incentivize the private sector to come up with innovative solutions in this field. One such circular system in the United States could involve the use of food waste generated by various branches of the Department of Defense (DOD) as feedstock for producing biofuel. The energy produced from the process, whether electricity or liquid biofuel, can be used to either power DOD facilities or to fuel non-combat DOD vehicles.

Similar to the public transportation program in Finland that runs on locally produced bio-gas,[142] DOD could create a new ecosystem to utilize the food waste generated by various bases. An in-depth market research prior to the procurement would be required detailing the implementation challenges. Factors such as allocation of risk, cost-gap analysis, local conditions and markets,

 DoD already has identified food waste as a major problem. It is the main component of waste streams in large military installations. DoD has been exploring opportunities to divert food waste from landfills by using dehydrators to separate the solid waste from the liquid waste and discarding the liquid waste into the sewage system. This leads to significant space savings in landfills, however it does not reduce waste or create energy. DoD is also looking at ways to prevent food waste in the first place.

---

[140] Alex Maag, *supra note 133*, at 3 (listing the food items used: American Cheese, Canned Chicken, Instant Potatoes, Green Beans, White Rice, Apple Sauce and Butter).
[141] Sharon Gaudin, *supra note 138* (stating that 'red mud' which is an inexpensive, stable and reliable waste created as a byproduct of aluminum production, is being investigated as a potential catalyst).
[142] *See* Katrina Alhola, *supra note 18*, at 139.

DoD has conducted studies in the viability of renewable energy production from its installations' solid waste through anaerobic digestion methods. However, as the WPI research showed, there is a vast potential for innovation that has not yet been realized. For instance, DoD has not explored the viability of coupling hydrothermal liquefaction and anaerobic digestion. Anaerobic digestion of post-hydrothermal liquefaction wastewater has the potential for improved energy efficiency than either process alone, especially with the additional efficiencies that the WPI project was able to create. Another potential avenue of innovation could be to improve the biodegradability of post-hydrothermal liquefaction wastewater. Studies partly funded by the Chinese government have shown that this could be achieved through ozone treatment. Coupling the two process, while achieving efficiencies in both through innovation could solve the major problem of food waste on large DoD bases and installations, while creating energy offsets and carbon emission savings. As this paper tried to show in Part I, applying circular economy practices to procurement could be the solution that spurs innovation required to solve this problem.

Achieving commercial viability of producing biofuel from food waste feedstock is not likely in the short term without significant innovation. Therefore, it is most likely that any biofuel created by DoD using this methodology in the short term will be used for government purposes to offset some of DoD's vast energy demands. GATT Article III 8(a) also provides an exception from National Treatment requirements for products purchased for government purposes. Thus, DoD's purchase of biofuel produced through this program with a price premium, would not be a violation of National Treatment principle under the GATT.

In the United States, as pointed out earlier in the paper, biofuel is considered an agricultural product due to it being 'bio based.' This means any bio fuel produced in the United States is outside the coverage of the GPA as well, and hence can be protected against foreign competition without violating the non-discrimination principles of the GPA.

Even if biofuel made from food waste is not considered an agricultural product in the future due to a feedstock based certification process, DoD will be able to procure such biofuel from its own facilities or program. Including technical specifications such as fuel generating domestic carbon savings or those focusing on food waste being reduced in DoD facilities will not be in violation of non-discrimination principles of the GPA. Article X(6) of the GPA allows parties to include technical specifications in their procurements to promote their environmental interests. It states, "a Party, including its procuring entities, may, in accordance with this Article, prepare, adopt or apply technical specifications to promote the conservation of natural resources or protect the environment."[143] Furthermore, Award Criteria in Article XIII(4)(a) of the GPA also allows parties to make awards based on which bidder represents the most overall value to the government, allowing them to consider environmental factors while making an award. Thereby further protecting any domestic biofuel programs that the government may invest in to promote innovation. Provisions of the revised GPA can thus be seen as reflective of the member states' wishes to use it as a tool to promote sustainable development.

---

[143] *Id.*